\documentclass[amsmath,amssymb,showkeys, showpacs]{revtex4}

\begin{document}
\hspace*{5 in}CUQM - 145
\vskip 0.4 in

\title{Exact and approximate solutions
of Schr\"odinger's equation for \\a class of trigonometric potentials}
\author{Hakan Ciftci$^{\ast }$, Richard L Hall$^\dagger$ and Nasser Saad$^\ddagger$}
\email{hciftci@gazi.edu.tr, rhall@mathstat.concordia.ca, nsaad@upei.ca}
\affiliation{$^{\ast }$Gazi \"{U}niversitesi, Fen-Edebiyat Fak\"{u}ltesi, Fizik B\"{o}l%
\"{u}m\"{u}, 06500 Teknikokullar-Ankara, Turkey.\\
$^\dagger$Department of Mathematics and Statistics, Concordia University,
1455 de Maisonneuve Boulevard West, Montr\'eal,
Qu\'ebec, Canada H3G 1M8\\
$^\ddagger$Department of Mathematics and Statistics,
University of Prince Edward Island,
550 University Avenue, Charlottetown,
PEI, Canada C1A 4P3.}
\begin{abstract}
\noindent The asymptotic
iteration method is used to find exact and approximate solutions of Schr\"odinger's equation for a number of one-dimensional trigonometric potentials (sine-squared, double-cosine, tangent-squared, and complex cotangent). Analytic and approximate solutions
are obtained by first using a coordinate transformation to reduce the Schr\"odinger equation to a  second-order differential equation with an appropriate form. The asymptotic
iteration method is also employed indirectly to obtain the terms in perturbation expansions, both for the energies and for the corresponding eigenfunctions.
\end{abstract}
\pacs{31.15.-p 31.10.+z 36.10.Ee 36.20.Kd 03.65.Ge.\\ }
\keywords{Trigonometric potentials, Sine-squared potential,  Double-cosine potential, Tangent-squared potential,  Cotangent complex potential, Asymptotic
iteration method, Perturbation method.}
\maketitle

\section{Introduction}

\noindent Analytic solutions of Schr\"odinger's equation are possible only for few potential models. Although there are a number of approximation methods available in the literature  to estimate the solution of Schr\"odinger's equation, they cannot usually be used to study both the exact solutions as well as approximate solutions. The asymptotic iteration method was recently introduced and this can indeed find both the exact and numerical solutions of Schr\"odinger equation for a very large number of quantum potentials \cite{aim1}.  If the eigenvalue problem is analytically solvable, the aysmptotic iteration method provides explicit algebraic expressions for the  energy eigenvalues and the corresponding wave functions \cite{aim2}. Meanwhile, if the eigenvalue problem is not exactly solvable, the asymptotic iteration method provides approximate solutions (for the eigenvalues and eigenfunctions) through its termination condition \cite{aim3}. 
\vskip0.1true in
\noindent In the present work we use the asymptotic iteration method (AIM) to provide both exact and approximate solutions of Schr\"odinger's equation for a class of one-dimensional trigonometric potentials. The approach discussed here can be easily applied  to solve Schr\"odinger's equation for other potentials, such as the Rosen-Morse potential, the Eckart potential, and the P\"oschl-Teller potential.
\vskip0.1true in
\noindent Another purpose of the present work is to explore  perturbation theory based on the asymptotic iteration method to obtain a perturbation expansion for the energy eigenvalues \cite{aim5,bakarat}, up to the fourth-order of approximation in some cases.  The advantage of the perturbation method presented here lies in the direct computation of the coefficients in the perturbation series without the necessity of having the eigenfunctions of the unperturbed problem available.
\vskip0.1true in

\noindent A variety of applications of the asymptotic iteration method are studied in the present work: first as a method for obtaining exact analytic solutions, second as a direct approximation method, thirdly as a method for obtaining  the terms in perturbation series. The organization of the paper is as follows: in the next section we present a brief introduction to the asymptotic iteration method; this is followed by the description of its perturbation version in section III.  In section IV, we study the sine-square potential defined by
\begin{equation}\label{eq1}
{V_1(x)=}\left\{ 
\begin{array}{ll}
V_{0}\sin ^{2}(\frac{x}{a}),&x\in \left( -\frac{\pi a}{2},
\frac{\pi a}{2}\right)  \\ 
\infty,&\left\vert x\right\vert >\frac{\pi a}{2}%
\end{array}%
\right. 
\end{equation}%
We use the standard asymptotic iteration method to obtain accurate numerical approximations to the energy eigenvalues. We then analyze the eigenvalue problem using our perturbation method, regarding $V_1(x)$ as a perturbation of the zero potential. In section V, we study the double-cosine potential defined by
\begin{equation}\label{eq2}
{V_2(x)=}\left\{ 
\begin{array}{ll}
V_1\cos(x)+V_2\cos(2x),&0<x
<2\pi  \\ 
\infty,& x<0 ~and ~ x>2\pi%
\end{array}%
\right. 
\end{equation}%
by means of the asymptotic iteration method. In section VI, we apply the asymptotic iteration method to obtain exact analytical solutions for the tangent potential
\begin{equation}\label{eq3}
{V_3(x)=}V_{0}\tan ^{2}(\frac{x}{a})\text{, }x\in \left( -\frac{\pi a}{2}\text{%
,}\frac{\pi a}{2}\right). 
\end{equation}%
Finally in section VII, we investigate the exact solutions of the complex cotangent potential 
\begin{equation}\label{eq4}
{V_4(x)=}\left\{ 
\begin{array}{ll}
i\nu \cot (\frac{x}{a}),&0<x<a\pi  \\ 
\infty,&x=0\text{ and }x=a\pi 
\end{array}%
\right. 
\end{equation}
\section{A brief introduction to the asymptotic iteration method}
\noindent A detailed description of the asymptotic iteration method can be found in Refs.~\cite{aim1,aim2,aim3,aim4,aim5,aim6}. For completeness, we present here a brief outline of the method. Given  sufficiently differentiable functions $\lambda _{0}(x)$ and $s_{0}(x)$, the second-order linear differential equation
\begin{equation}\label{eq5}
y^{\prime \prime }(x)=\lambda _{0}(x)y^{\prime }(x)+s_{0}(x)y(x)
\end{equation}
has a general solution 
\begin{equation}\label{eq6}
y(x)=\exp\left(-\int^{x}\alpha dt\right)\left( C_{2}+C_{1}\int^{x}\exp \left(
\int^{t}(\lambda _{0}(\tau )+2\alpha (\tau ))d\tau \right) dt\right)
\end{equation}%
where $C_1$ and $C_2$ are constants, if for sufficiently large $n>0$,
\begin{equation}\label{eq7}
{\frac{s_{n}}{\lambda _{n}}}={\frac{s_{n-1}}{\lambda _{n-1}}}\equiv \alpha
\end{equation}
where
\begin{equation}\label{eq8}
\lambda _{n}=\lambda _{n-1}^{\prime }+s_{n-1}+\lambda _{0}\lambda
_{n-1},~~~~ s_{n}=s_{n-1}^{\prime }+s_{0}\lambda _{n-1},\qquad n=1,2,\dots.
\end{equation}
Note that the termination condition (\ref{eq7}) can be written equivalently as
\begin{equation}\label{eq9}
{\delta_{n} (x)=}\lambda _{n}(x)s_{n-1}(x)-s_{n}(x)\lambda _{n-1}(x)=0,
\end{equation}
\noindent where $n$ refers to the iteration number. The $\delta_n(x)$ function has a crucial role for solving the boundary-value problem:
since the coefficients $s_{n}$ and $\lambda _{n}$ depend on the (unknown)
eigenvalue $E$ that are usually determined by the roots of equation (\ref{eq9}). If, for
$\lambda_0$ and $s_0$ and exact $E$, the termination condition $\delta_n(x)\equiv 0$ is satisfied at every $x$ point,
then the problem is called `exactly solvable' and in this case the radial quantum number (or the number of nodes in the
wave function) is
equal to the iteration number. If the problem is not analytically
solvable, the expression for $\delta_n(x)$ depends on both $x$ and the unknown $E$ where in this
case the eigenvalues are obtained through the numerical roots of (\ref{eq9}) at some suitable (chosen) $x_{0}$-point. This initial value may be determined generally as the maximum value of the asymptotic wave function or the
minimum value of the potential \cite{aim3}, however, the initial value is easily determined if the spacial domain for the given eigenvalue problem is bounded. Indeed, for this class of problem where $x\in [a,b]$, the average point $x_0=(a+b)/2$ will be sufficient to initiate the iteration process. 
\section{A perturbation theory via asymptotic iteration method}
\noindent In Ref. \cite{aim5}, the present authors gave a new perturbation theory based on the asymptotic iteration method \cite{aim5}. In order to apply the method for a class of trigonometric potentials regarded as perturbations of the zero potential, we outline our perturbation approach. Assume the potential is given as follows
\begin{equation}\label{eq10}
V(x)=V_{0}(x)+\mu~ V_{1}(x)
\end{equation}
where $V_0(x)$ is a solvable potential, $V_1(x)$ is the perturbation potential, and $\mu$ is a perturbation
expansion parameter. 
The one-dimensional Schr\"odinger equation reads%
\begin{equation}\label{eq11}
-\frac{d^{2}\Psi }{dx^{2}}+\left( V_{0}(x)+\mu V_{1}(x)\right) \Psi
(x)=E\Psi (x).
\end{equation}%
Substitute $\Psi(x)= y(x)f(x)$ where $y(x)$ is an asymptotic solution satisfying the boundary conditions for  the eigenvalue problem (\ref{eq11}). Substitution yields the following differential equation for $f(x)$
\begin{equation}\label{eq12}
f^{\prime \prime }(x)=\lambda _{0}(x,\mu )f^{\prime }(x)+s_{0}(x,\mu
)f(x),
\end{equation}%
which is in a form suitable for the asymptotic iteration method. Using (\ref{eq12}), we can now compute the sequences $\lambda_n(x,\mu)$ and $s_n(x, \mu)$.  The termination condition $\delta_n(x,\mu)$, as given by (\ref{eq12}), can be written as
\begin{equation}\label{eq13}
\delta_n (x; \mu,E )=\lambda_{n}(x;\mu,E)s_{n-1}(x;\mu,E)-s_{n}(x;\mu,E)\lambda_{n-1}(x;\mu,E)=0.
\end{equation}%
Expanding $\delta_n(x;\mu,E)$ about $\mu =0$, we obtain the following infinite series
\begin{equation}\label{eq14}
{\delta_n(x; \mu, E )=}\sum_{k=0}^{\infty }{\delta }_{n,k}(x;0,E)\cdot \mu^{k}
\end{equation}%
that resembles Taylor's expansion with
\begin{equation}\label{eq15}
{\delta }_{n,k}{(x,0,E)}=\frac{1}{k!}\frac{\partial ^{k}{\delta_n (x;\mu,E )}}{%
\partial \mu^{k}}\bigg|_{\mu =0}=\frac{1}{k!}\frac{\partial ^{k}{\left[\lambda_{n}(x;\mu,E)s_{n-1}(x;\mu,E)-s_{n}(x;\mu,E)\lambda_{n-1}(x;\mu,E)\right]}}{%
\partial \mu^{k}}\bigg|_{\mu =0},~ k=0,1,2,\dots.
\end{equation}%
According to the standard theory of the asymptotic iteration method, the eigenvalues are obtained by the solution of the equation  $\delta_n(x;\mu, E)=0$ for arbitrary $\mu$ and thus the perturbation coefficients satisfy  
\begin{equation}\label{eq16}
\delta_{n,k}{(x;0,E)}=0.
\end{equation}%
It is clear from the standard theory of the asymptotic iteration method that the root of the equation $\delta_{n,0}(x;0,E)= 0$ gives the eigenvalue of the non-perturbative Hamiltonian, say, $E_{n,0}$ that may depend on $n$ (the number of iterations used by the asymptotic iteration method). Meanwhhile $\delta_{n,1}(x;0,E) = 0$
gives the first-order correction to the eigenvalue $E_{n,0}$, say $E_{n,1}$, $\delta_{n,2}(x;\mu_0,E) = 0$
gives the second-order correction to the eigenvalue $E_{n,0}$, say $E_{n,2}$, and so on. This procedure 
allows us to find the coefficients of the eigenvalue expansion 
\begin{equation}\label{eq17}
E=\sum_{k=0}^{\infty }E_{n,k}\cdot {\mu }^{k}
\end{equation}%
where (again) $E_{n,k}$ are computed by solving ${\delta }_{n,k}{(x,0,E)}=0$. It is an attractive
feature of the asymptotic iteration method that we can also find the eigenfunctions by means of equation (\ref{eq9}) where it is sufficient to compute 
\begin{equation}\label{eq18}
f(x)=\exp\left( -\int^{x}\alpha \left( t,\mu \right) dt\right),\qquad \alpha(x,\mu)\equiv {s_n(x,\mu)/\lambda_n(x,\mu)}
\end{equation}%
Again if we expand $\alpha(x,\mu)$ about $\mu=0$, we obtain the following
series
\begin{equation}\label{eq19}
\alpha (x,\lambda )=\sum_{k=0}^{\infty }{\alpha }_{k}{(x,0)\mu }^{k},
\end{equation}%
where ${\alpha }_{k}(x,0)$, $k=0,1,2,\dots$ is given by
\begin{equation}\label{eq20}
{\alpha }_{k}{(x,0)}=\frac{1}{k!}\frac{\partial ^{k}{\alpha (x,\lambda )}}{%
\partial \mu^{k}}\bigg|_{\mu =0}.
\end{equation}%
The correction of the wave function is then given by%
\begin{equation}\label{eq21}
f(x)=\exp\left(-\sum_{k=0}^\infty \mu^k\int^x \alpha_k(\tau,0)d\tau\right)=  \prod\limits_{k=0}^{\infty }f_{k}(x),~~
f_{k}(x)=\exp\left( -\mu^k\int^{x}\alpha _{k}\left( \tau,0\right) d\tau\right),~~k=0,1,2,\dots.
\end{equation}
Thus we have a perturbation method  (based on the asymptotic iteration method) that could in principle provide a complete solution to the given eigenvalue problem.
\section{Sine-squared potential}

\noindent In this section we present the solution of the one-dimensional Schr\"odinger equation for
the  trigonometric sine-square potential (\ref{eq1})
\begin{equation}\label{eq22}
\left[ -\frac{d^{2}}{dx^{2}}+V_{0}\sin ^{2}\left(\frac{x}{a}\right)\right] \Psi 
(x)=E\Psi (x),\qquad {\Psi }\left( \frac{\pi a}{2}%
\right) = {\Psi }\left( -\frac{\pi a}{2}%
\right)=0,\qquad a\neq 0.
\end{equation}
The behavior of the wave function at the boundaries suggests a solution of the form
\begin{equation}\label{eq23}
\Psi (x)=\cos \left(\frac{x}{a}\right)f(x).
\end{equation}%
Substituting for $\Psi(x)$ in equation (\ref{eq22}), we obtain%
\begin{equation}\label{eq24}
f^{\prime \prime }(x)=\frac{2}{a}\tan \left(\frac{x}{a}\right)f^{\prime }(x)+\left( 
\frac{1}{a^{2}}-E+V_{0}\sin ^{2}\left(\frac{x}{a}\right)\right) f(x).
\end{equation}%
A further change of the independent variable $x=a\arcsin(y)$ yields 
\begin{equation}\label{eq25}
f^{\prime \prime }(y)=\frac{3y}{1-y^{2}}f^{\prime }(y)+\left( \frac{\omega }{%
1-y^{2}}-\mu\right) f(y)
\end{equation}%
where $\omega =1+\mu -a^{2}E$ and $\mu =V_{0}a^{2}$. We now find solutions of this differential equation by using two different  approaches: (1) the standard asymptotic iteration method discussed in section II; (2) the perturbation method developed in section III.

\subsection{The standard asymptotic iteration method}

\noindent We may initiate the iteration method with 
$$
\left\{ \begin{array}{l}
\lambda_0= \frac{3y}{1-y^{2}} \\ \\
 s_0= \frac{1+V_{0}a^{2}-a^{2}E }{%
1-y^{2}}-V_{0}a^{2}. 
       \end{array} \right.
$$
The sequences $s_n$ and $\lambda_n$, $n = 1, 2, \dots$  can then be evaluated using (\ref{eq8}). These quantities, for fixed $a$ and $V_0$, are functions of the parameter $E$ and the variable $y$. For a suitable initial value of $y=y_0<1$, the eigenvalues $E$ are computed using the roots of the termination condition (\ref{eq12}). We select the initial value $y_{0}=0$ that locate potential minimum as given by (\ref{eq1}). In Table 1, we report our numerical calculation of $E_n$ for different values of the parameter $\mu$, where $n$ represent the node of the wave function solution, as given using the asymptotic iteration method.
\begin{table}[!h]
\caption{The computation of $a^2E$ where $\mu=V_0a^2$ using the standard asymptotic iteration method. The subscript refer to the number of iteration.\\ }
\centering
\begin{tabular}{|l|l|l|l|l|l|l|}
\hline
$n$ & $a^{2}E(\mu =0)$ & $a^{2}E(\mu =0.1)$ &$a^{2}E(\mu =0.5)$ &$a^{2}E(\mu =1)$ & $a^{2}E(\mu =5)$&
$a^{2}E(\mu =10)$ \\ \hline
$0$ & $1$ & $1.024~922~118~9_{10}$& $1.123~077~224~8_{14} $ & $1.242~428~826~0_{14}$ &$2.082~985~293~2_{20}$& $2.923~668~494~2_{22}$ \\ \hline
$1$ & $4$  &$4.049~947~916~8_{11}$ &$ 4.248~698~005~0_{14}$& $4.494~793~078~6_{15}$ &$6.370~661~125~0_{21}$ & $8.492~474~366~7_{23}$ \\ \hline
$2$ & $9$ & $9.050~038~818~6_{12}$&$9.250~946~208~9_{16}$ & $9.503~664~867~0_{16}$ & $11.569~339~156~9_{22}$& $14.185~709~970~1_{24}$ \\ \hline
$3$ & $16$ & $16.050~020~833~2_{13}$ & $ 16.250~520~743~8_{17}$& $16.502~081~901~0_{17}$ & $18.551~201~398~4_{23}$&$21.194~837~346~9_{27}$ \\ \hline
$4$ & $25$ & $25.050~013~020~8_{14}$ &$ 25.250~325~524~1_{16}$ &$25.501~302~132~2_{18}$&$ 27.532~566~336~_{24}1$ & $30.130~124~200~2_{28}$  \\ \hline
$5$ & $36$ & $36.050~008~928~6_{15}$ &$36.250~223~215~3_{17}$& $36.500~892~873~8_{21}$ &$ 38.522~331~587~4$& $41.089~436~879~7_{27}$ \\ \hline
\end{tabular}
\label{table:Tab1} 
\end{table}
\subsection{Perturbation theory via the asymptotic iteration method}
\noindent The exact values of $a^2E$ in the case $\mu=0$ as shown in Table 1 do not come as a surprise for we  have noted the problem is equivalent to solving Schr\"odinger equation for the particle in rigid box $V_0=0$; namely for the potential
\begin{equation}\label{eq26}
{V(x)=}\left\{ 
\begin{array}{ll}
0,& x\in \left( -\frac{\pi a}{2}\text{,}%
\frac{\pi a}{2}\right)  \\  
\infty &\left\vert x\right\vert >\frac{\pi a}{2}.%
\end{array}%
\right. 
\end{equation}%
The differential equation (\ref{eq25}) reads for $\mu=0$
\begin{equation}\label{eq27}
(1-y^2)f^{\prime \prime }(y)-{3y} f^{\prime }(y)-\omega f(y)=0
\end{equation}%
 can immediately solve with polynomial solutions are given for
\begin{equation}\label{eq28}
\omega=-n(n+2)\Rightarrow a^2 E=(n+1)^2
\end{equation}
by
\begin{equation}\label{eq29}
f_n(y)={}_2F_1(-n,n+2;{3\over 2};{1-y\over 2})\Rightarrow f_n(x)={}_2F_1(-n,n+2;{3\over 2};{1\over 2}- {1\over 2}{\sin({x\over a})}).
\end{equation}
\subsubsection{Energy Expansion}
\noindent Suppose that the ${a}^{2}E$ can be written as
follows%
\begin{equation}\label{eq30}
{a}^{2}E_n=\sum_{k=0}^{\infty }\nu _{n,k}\mu^{k}=\nu_{n,0}+\nu_{n,1}\cdot \mu+\nu_{n,2}\cdot \mu^2+\dots
\end{equation}%
In order to get a zeroth-order approximation to ${a}^{2}E$ using the iteration method discussed in section 3, we have for
\begin{equation}\label{eq31}
{\delta }_{0}{(x,0)}=0
\end{equation}%
that exact values 
\begin{equation}\label{eq32}
\nu _{n,0}=(n+1)^{2},\qquad n=0,1,2,...
\end{equation}%
as expected using the exact solutions mentioned earlier. Thus for the ground state $n=0$, we have
\begin{equation}\label{eq33}
{a}^{2}E_0=1+\nu_{0,1}\cdot \mu+\nu_{0,2}\cdot \mu^2+\dots
\end{equation}%
In order to find the first order approximation $\nu_{0,1}$, we take ${a}^{2}E_0=1+\nu _{0,1}\mu$ and solve the
termination equation 
\begin{equation}\label{eq34}
{\delta }_{1}(y,0)={\partial\delta(y,\mu)\over \partial \mu}\bigg|_{\mu=0}=0
\end{equation}%
where $\delta(y,\mu)$ is now computed using
$$
\left\{ \begin{array}{l}
\lambda_0= \frac{3y}{1-y^{2}} \\ 
 s_0=\frac{-\nu _{0,1}\mu+\mu y^2}{%
1-y^{2}}. 
       \end{array} \right.
$$
Straightforward computation using the sequences $\lambda_n$ and $s_n$, we obtain for the second-iteration the  following expression for $\delta(y,\mu)$
$$
\delta(y,\mu)={-\mu(-\mu^2 y^6+9\mu y^4+3\mu^2\nu_{0,1}y^4+6\mu y^2-6\mu\nu_{0,1} y^2-3\mu^2\nu_{0,1}^2y^2-6+2\nu_{0,1}\mu+\nu_{0,1}^3\mu^2+24\nu_{0,1}-11\nu_{0,1}^2\mu)\over (y^2-1)^3}.
$$
Thus
\begin{equation}\label{eq35}
{\delta }_{1}(y,0)={\partial\delta(y,\mu)\over \partial \mu}\bigg|_{\mu=0}=24\nu_{0,1}-6=0\Rightarrow \nu_{0,1}={1\over 4}
\end{equation}
as the first correction for the eigenvalue. A similar computation using $a^2E_0=1+{1\over 4}\mu+\nu_{0,2}\mu^2$ shows that
$\nu _{0,2}=-{1/128}$, and, using $a^2E_0=1+{(1/4)}\mu-{(1/128)}\mu^2+\nu_{0,3}\mu^3$, that $\nu _{0,3}={1/4096}$, and for $a^2E_0=1+{(1/4)}\mu-{(1/128)}\mu^2+{(1/4096)}\mu^3+\nu_{0,4}\mu^4$ that $\nu _{0,4}=-{1/393216}$, where in each case we update the expressions for $\lambda_0$ and $s_0$ to initiate the computation of $\delta(y,\mu)$.
Thus the ground-state energy eigenvalue expansion takes the form
\begin{equation}\label{eq36}
a^2E_{0}=1+{1\over 4}\mu-{1\over 128}\mu^2+{1\over 4096}\mu^3-{1\over 393216}\mu^4+\dots,
\end{equation} 
for sufficiently small $\mu$ as usual for standard perturbation theory. For the first-excited state, we obtain the eigenvalue expansion
\begin{equation}\label{eq37}
a^2E_{1}=4+{1\over 2}\mu-{1\over 192}\mu^2+\dots,
\end{equation} 
and for arbitrary $n\geq 2$,
\begin{equation}\label{eq38}
a^2E_{n}=(n+1)^{2}+\frac{\mu}{2}+\frac{\mu^{2}}{32n\left( n+2\right) }+\dots.
\end{equation}

\begin{table}[!h]
\caption{The computation of $a^2E$ for given $\mu=V_0a^2$ by perturbation theory via the asymptotic iteration method.\\ }

\centering
\begin{tabular}{|l|l|l|l|}
\hline
$n$ & $a^{2}E(\mu =0.1)$ &$a^{2}E(\mu =0.5)$ &$a^{2}E(\mu =1)$ \\ \hline
$0$ & $1.024~922~118~9$& $1.123~077~233~6$ & $1.242~429~097~5$\\  \hline
$1$ &$4.049~947~916~7$ &$4.248~697~916~7$& $4.494~791~666~7$ \\ \hline
$2$ & $9.050~039~062~5$&$9.250~976~562~5$ & $9.503~906~250$  \\ \hline
\end{tabular}
\label{table:Tab2} 
\end{table}
\subsubsection{Wave function expansion}
\noindent In order to calculate the perturbation expansion for the wave function, we note first that the zeroth-order
solution of the Eq.(\ref{eq22}), using equation (\ref{eq29}), is%
\begin{equation}\label{eq39}
f_n(y)={}_2F_1(-n,n+2;{3\over 2};{1-y\over 2}),\qquad n=0,1,2,\dots.
\end{equation}
Unlike the standard perturbation theory, the first-order corrected wave function using our approach can be computed without the use of  the zeroth-order
solution. Indeed, from (\ref{eq20}) we have for the first correction, using $a^2E=1+\nu_{0,1}\mu$, 
\begin{equation}\label{eq40}
\alpha(y,\mu)=-{\mu y(-2+5\nu_{0,1}-3y^2)\over 3+12y^2+\mu y^2-\nu_{0,1} \mu+\mu\nu_{0,1} y^2-\mu y^4}
\end{equation}
Thus
\begin{equation}\label{eq41}
{\partial \alpha(y,\mu)\over \partial \mu}\bigg|_{\mu=0}=-{y(-2+5\nu_{0,1}-3y^2)\over 3+12y^2}
\end{equation}
and 
\begin{equation}\label{eq42}
\int^y {\partial \alpha(\tau ,\mu)\over \partial \mu}\bigg|_{\mu=0}d\tau ={y^2\over 8}+{5\over 96}\ln(1+4y^2)-{5\nu_{0,1}\over 24}\ln(1+4y^2).
\end{equation}
However, since $\nu_{0,1}={1\over 4}$ we have
\begin{equation}\label{eq43}
\int^y {\partial \alpha(\tau ,\mu)\over \partial \mu}\bigg|_{\mu=0}d\tau ={y^2\over 8}.
\end{equation}
Finally, the first-order approximation for the wave function is
\begin{equation}\label{eq44}
f_1(y)=\exp(-{\mu\over 8}y^2)\Longrightarrow \Psi_0(x)\sim \cos\left({x\over a}\right)e^{-{\mu\over 8}\sin^2\left({x\over a}\right)},
\end{equation}
where $\psi_0(x)$ is the first-order approximation of the ground-state wave function solution of (\ref{eq23}). Similarly, we can compute the first-order approximation for $n=1$ where we have
 \begin{equation}\label{eq45}
f_1(y)=\exp(-{\mu\over 12}y^2).
\end{equation}
the first-order approximation for $n=2$, we have
 \begin{equation}\label{eq46}
f_1(y)=\exp\left(\left(-{y^2\over 16}+{3\over 64(4y^2-1)}\right)\mu\right).
\end{equation}
Finally, for the first-order approximation in the case of $n=3$, we have
 \begin{equation}\label{eq47}
f_1(y)=\exp\left(\left(-{y^2\over 20}+{1\over 60(2y^2-1)}\right)\mu\right).
\end{equation}
\subsection{Equivalent potentials}
\noindent We may note using the trigonometric identities
\begin{equation}\label{eq48}
\sin^2(\theta)=1-\cos^2(\theta)\qquad\mbox{and}\qquad \sin^2(\theta)={1\over 2}\left(1-\cos(2\theta)\right),\qquad \sin^2(\theta)=\cos^2(\theta)-\cos(2\theta)
\end{equation}
that the eigenvalue solutions of the following schr\"odinger equations for trigonometric potentials confined between two rigid walls of infinite potential energy
\begin{align}
\left[ -\frac{d^{2}}{dx^{2}}-V_{0}\cos ^{2}\left(\frac{x}{a}\right)\right] \Psi_\eta
(x)&=E_{\eta}\Psi_\eta (x),\qquad {\Psi_\eta }\left( x\rightarrow \mp \frac{\pi a}{2}%
\right) =0,\label{eq49}\\
\left[ -\frac{d^{2}}{dx^{2}}-{1\over 2}V_{0}\cos \left(\frac{2x}{a}\right)\right] \Psi_\xi
(x)&=E_{\xi}\Psi_\xi (x),\qquad {\Psi_\xi }\left( x\rightarrow \mp \frac{\pi a}{2}%
\right) =0,\label{eq50}\\
\left[ -\frac{d^{2}}{dx^{2}}+V_{0}\cos ^{2}\left(\frac{x}{a}\right)-V_0
\cos\left(\frac{2x}{a}\right)\right] \Psi_\zeta
(x)&=E_{\zeta}\Psi_\zeta (x),\qquad {\Psi_\zeta }\left( x\rightarrow \mp \frac{\pi a}{2}%
\right) =0,\label{eq51}
\end{align}
are related to the eigenvalues of the sine-square potential as given by equation (\ref{eq22}). These relations are easily derived and are given explicitly by
 \begin{align}\label{eq52}
E_{\eta}&=E-V_0,\qquad E_\xi=E-{1\over 2}V_0,\qquad
E_{\zeta}=E.
\end{align}
\section{double-cosine potential}
\noindent In this section, we consider a single particle in the double cosine potential \cite{Hai},
\begin{equation}\label{eq53}
{V_2(x)=}\left\{ 
\begin{array}{ll}
V_1\cos(x)+V_2\cos(2x),&0<x
<2\pi  \\ 
\infty,& x<0 ~and ~ x>2\pi%
\end{array}%
\right. 
\end{equation}%
and obeys the Schr\"odinger equation
\begin{equation}\label{eq54}
\left[ -\frac{d^{2}}{dx^{2}}+V_1\cos(x)+V_2\cos(2x)\right] \psi 
(x)=E\Psi (x),\qquad {\psi }\left( 0%
\right) = {\psi }\left(2\pi%
\right)=0.
\end{equation}
The maximum of the potential (\ref{eq53}) occurs within the given interval between $x=0$ and $x=2\pi$ and has the value of $V_{max}=V_1+V_2$, while the minimum occurs at  
$x=\pi$ with the value of  $V_{min}=V_2-V_1.$ Clearly, if $V_1=0$ and $V_2>0$, the problem corresponds to Mathieu's equation \cite{mclachlan}. Assume the general solution of the differential equation (\ref{eq54}) that satisfies the boundary conditions takes the form
$\psi(x)=\sin\left({x/2}\right)f(x)$. Then direct substitution, using equation \eqref{eq54}, yields the second order differential equation for $f(x)$ as
\begin{equation}\label{eq55}
f''(x)=-\cot\left({x\over 2}\right)f'(x)+({1\over 4}+V_1\cos(x)+2V_2\cos^2(x)-V_2-E)f(x)
\end{equation}
Assuming, further, that 
\begin{equation}\label{eq56}
f(x)=\exp\left(-{V_1\over 2}\cos({x})\right)g(x),
\end{equation}
we obtain
\begin{equation}\label{eq57}
g''(x)=-\left(\cot\left({x\over 2}\right)+V_1\sin\left(x\right)\right)g'(x)+\left({1\over 4}- {V_1\over 2}-{V_1^2\over 4}+({V_1^2\over 4}+2V_2)\cos^2(x)-V_2-E\right)g(x)
\end{equation}
where we have used  the trigonometric identity $\cos(x)=-1+2\cos^2(x/2)$. Finally with the substitution 
$y=\cos({x/2})$ we obtain
\begin{align}\label{eq58}
g''(y)&=-\left({y\over {1-y^2}}
+4V_1y\right)g'(y)+\left({1-2V_1+4V_2-4E\over 1-y^2}-4(V_1^2+8V_2)y^2\right)g(y)
\end{align}
Clearly we have the exact solution 
\begin{equation}\label{eq59}
g_0(x)=1,\qquad E_0={1\over 4}-{V_1\over 2}+V_2\qquad\text{subject to}\qquad
V_2=-{1\over 8}V_1^2
\end{equation}
that is
\begin{equation}\label{eq60}
g_0(x)=1,\qquad E_0={1\over 4}-{V_1\over 2}-{1\over 8}V_1^2.
\end{equation}
For excited states we have applied the asymptotic iteration method that we used with 
\begin{equation}\label{eq61}
\left\{ \begin{array}{l}
 \lambda_0=-{y\over {1-y^2}}
+4V_1y  \\ \\
s_0={1-2V_1+4V_2-4E\over 1-y^2}-4(V_1^2+8V_2)y^2.
       \end{array} \right.
 \end{equation}      
 and a starting value of $y_0=0$ corresponding to the minimum of the double-cosine potential at $x_0=\pi$.   In Table \ref{table:Tab3}  we report the first few excited states along with the iteration number used by the asymptotic iteration method. All the eigenvalues are accurate for the given number of decimals shown and can easily be obtained  with any desired precision.

\begin{table}[!h]
\caption{The eigenenergies of Schr\"odinger's equation with the double cosine potential $V(x)=\cos(x)-{1\over 8}\cos(2x)$, where $y_0=0$.\\ }
\centering
\begin{tabular}{|l|c|l|c|}
\hline
$n$ & $E_n$&$n$ & $E_n$\\ \hline
$0$ & $-~0.375~000~000~000_3$&$5$&$6.270~895~381~051_{52}$\\  \hline
$1$ & $~~~0.625~000~000~000_{50}$&$6$& $9.014~386~533~890_{55}$\\ \hline
$2$ & $~~~1.258~305~195~063_{51}$&$7$&$12.260~510~491~188_{56}$\\ \hline
$3$ & $~~~2.311~907~935~564_{52}$&$8$&$16.008~020~136~795_{57}$\\ \hline
$4$ & $~~~4.031~642~235~192_{53}$&$9$&$20.256~322~480~092_{58}$\\ \hline
\end{tabular}
\label{table:Tab3} 
\end{table}

\noindent This  work can be easily extended to study the Schr\"odinger equation with the periodic potential
\begin{equation}\label{eq62}
\left[ -\frac{d^{2}}{dx^{2}}+v_1\cos\left({x\over k}\right)+v_2\cos\left({2x\over k}\right)\right] \psi 
(x)={\cal E}_{\{v_1,v_2\}}^k \psi (x),\quad 0<x<2k\pi,~~ k=1,2,\dots,\quad {\psi }\left( 0%
\right) = {\psi }\left(2k\pi%
\right)=0,
\end{equation}
with rigid walls at the boundaries. A simple scaling argument $x\rightarrow \sigma x$ with $\sigma=k$ shows that
\begin{equation}\label{eq63}
\left[ -\frac{d^{2}}{dx^{2}}+k^2v _1\cos({x})+k^2v_2\cos({2x})\right] \psi 
=k^2{\cal E}\psi,\quad 0<x<2\pi,~~ k=1,2,\dots,\quad {\Psi }\left( 0%
\right) = {\Psi }\left(2\pi%
\right)=0,
\end{equation}
so that we can easily compare with our earlier work concerning equation \eqref{eq54}, where in this case
\begin{equation}\label{eq64}
E(k^2v_1,k^2v_2)=k^2 {\cal E}_{\{v_1,v_2\}}^k.
\end{equation}
\section{Tangent-squared potential}
\noindent Recently, Ta\c{s}eli \cite{tac} studied the squared tangent potential $V(x)=\nu(\nu-1)\tan^2(x),~x\in (-\pi/2,\pi/2)$ also known as the symmetric P\"oschl-Teller potential \cite{nieto, gori}. It was  noted that the energy eigenvalues, but not the eigenfunctions, of the Hamiltonian
with the squared tangent potential on the symmetric interval $(-\pi/2, \pi/2)$ are precisely the same as
those of the Hamiltonian with the squared cotangent potential $V(x)=\nu(\nu-1)\cot^2 x$ on the asymmetric interval $(0, \pi)$ considered by
Marmorino in \cite{mar} (see also \cite{ma}). This observation can be easily verified  using the identity
\begin{equation}\label{eq65}
\tan(x+{\pi\over 2})=-\cot(x).
\end{equation}
In this section, we will used the asymptotic iteration method to study the Schr\"odinger equation with the confining potential
\begin{equation*}
{V(x)=}V_{0}\tan ^{2}\left(\frac{x}{a}\right)\text{, }\qquad x\in \left( -\frac{\pi a}{2}\text{%
,}~\frac{\pi a}{2}\right) ,\qquad V_0\geq 0
\end{equation*}%
contained within an infinite square well with sides at $x=\pm~ {\pi a}/{2}$ and $V_0$ gives an indication of how rapidly the potential increases within the well. In this case, the time-independent Schr\"odinger equation reads
\begin{equation}\label{eq66}
\left[ -\frac{d^{2}}{dx^{2}}+V_{0}\tan ^{2}\left(\frac{x}{a}\right)\right] \Psi
(x)=E\Psi (x),\qquad {\Psi }\left(-\frac{\pi a}{2}\right) =\Psi\left(\frac{\pi a}{2}\right)=0.
\end{equation}%
We, however, take a slightly different approach to solve \eqref{eq66} by letting the wave function assume the form
\begin{equation}\label{eq67}
\Psi (x)=\cos\left(\frac{x}{a}\right)\cdot f(x),
\end{equation}%
that satisfies the boundary conditions of \eqref{eq66}.
Again it is straightforward to verify that the substitution of (\ref{eq67}) into (\ref{eq66}), yields a differential equation for $f(x)$ as
\begin{equation}\label{eq68}
f^{\prime \prime }(x)=\frac{2}{a}\tan (\frac{x}{a})f^{\prime }(x)+\left( 
\frac{1}{a^{2}}-E+V_{0}\tan ^{2}(\frac{x}{a})\right) f(x).
\end{equation}%
A further change of variable $x=a\arcsin(y)$ implies%
\begin{equation}\label{eq69}
f^{\prime \prime }(y)=\frac{3y}{1-y^{2}}f^{\prime }(y)+\left( \frac{\omega }{%
1-y^{2}}+\frac{\mu y^{2}}{(1-y^{2})^{2}}\right) f(y),
\end{equation}%
and, with the substitution 
\begin{equation}\label{eq70}
f(y)=(1-y^{2})^{\alpha }g(y),
\end{equation}%
we finally obtain for $g(y)$ 
\begin{equation}\label{eq71}
g^{\prime \prime }(y)={{\left( 4\alpha +3\right) y}\over (1-y^2)}~g^{\prime
}(y)+{({\omega +2\alpha })\over (1-y^2)}~g(y),\quad y\in (-1,1).
\end{equation}%
where we set $\omega =1-a^{2}E$, $\mu =a^{2}V_{0}$ and 
$\alpha =\left(\sqrt{\mu +1}-1\right)/2.$ This differential equation is an example of an exactly solvable problem solved by means of the asymptotic iteration method. Indeed, with $$\lambda_0(y)={\left( 4\alpha +3\right) y\over 1-y^2}\qquad\text{and}\qquad s_0(y)={{\omega +2\alpha }\over 1-y^2},$$
the termination condition \eqref{eq9}, $\delta_n\equiv 0$ yields
\begin{align*}
\delta_1&=0\Longrightarrow (w+2\alpha)(w+6\alpha+3)=0\Longrightarrow w=-2\alpha, w=-6\alpha-3\\
\delta_2&=0\Longrightarrow (w+2\alpha)(w+6\alpha+3)(w+10\alpha+8)=0\Longrightarrow w=-2\alpha, w=-6\alpha-3, w=-10\alpha-8\\
\dots&=\dots
\end{align*}
and in general
\begin{equation*}
\delta_n=0\Longrightarrow \prod_{k=0}^n(w+k(k+4\alpha +2)+2\alpha)=0 \Longrightarrow w_n=-n(n+4\alpha +2)-2\alpha,\qquad n=0,1,2,\dots
\end{equation*}
Thus the energy eigenvalues are given by
\begin{equation}\label{eq72}
a^2E_n=n^2+(2n+1)(2\alpha+1),\qquad n=0,1,2,\dots
\end{equation}%
For the analytic solution of $g_n(y)$, we note by equation (6) that
\begin{align*}
n&=0\Longrightarrow g_0(y)=1\\
n&=1\Longrightarrow g_1(y)=y\\
n&=2\Longrightarrow g_2(y)=4(\alpha+1)y^2-1\\
n&=3\Longrightarrow g_3(y)=y(2(2\alpha+3)y^3-3)\\
n&=4\Longrightarrow g_4(y)=8(\alpha+2)(2\alpha+3)y^4-6(6+4\alpha)y^2+3\\
\dots&=\dots
\end{align*}
These polynomial solutions are easily generated by 
\begin{align}\label{eq73}
g_n(y)={}_2F_1(-n,n+4\alpha+2;
2\alpha+{3\over 2}, {1-y\over 2}),\qquad n=0,1,2,\dots
\end{align}
up to a multiplicative constant. Here, the Gauss hypergeometric function ${}_2F_1(\alpha,\beta;\gamma;x)$ is defined by \cite{andrews}
\begin{equation}\label{eq74}
{}_2F_1\left(\begin{array}{ll}
\alpha & \beta \\
\gamma & ~ \\
\end{array}\bigg|x\right)=\sum_{k=0}^\infty {(\alpha)_k(\beta)_k\over (\gamma)_k}{x^k\over k!}=1+{\alpha\beta\over \gamma}{x\over 1!}+{\alpha(\alpha+1)\beta(\beta+1)\over \gamma(\gamma+1)}{x^2\over 2!}+\dots
\end{equation}
and $(\lambda)_k$ denotes the Pochhammer symbol defined, in terms of Gamma functions, by
$$(\lambda)_k:= {\Gamma(\lambda+k)\over\Gamma(\lambda)}=\left\{ \begin{array}{ll}
 1 &\mbox{ if\quad $\left( k=0; \, \lambda \in {\Bbb C} \backslash \lbrace 0 \rbrace \right) $} \\
  \lambda(\lambda+1)(\lambda+2)\dots(\lambda+k-1) & \mbox{ if\quad $\left( k \in {\Bbb N}; \, \lambda \in {\Bbb C} \right).$}
       \end{array} \right.
$$ 
Thus the exact analytic solutions of Schr\"odinger's equation (\ref{eq66}) is
\begin{equation}\label{eq75}
\left\{ \begin{array}{l}
 \Psi_n(x)=A_n\cos^{2\alpha+1}\left({x\over a}\right){}_2F_1(-n,n+4\alpha+2;
2\alpha+{3\over 2}, {1\over 2}-{1\over 2}\sin\left({x\over a}\right)),\qquad n=0,1,2,\dots\\
\\
a^2E_n=n^2+(2n+1)(2\alpha+1),\qquad \alpha={1\over 2}\left(\sqrt{a^2V_0 +1}-1\right).
       \end{array} \right.
\end{equation}
where $A_n$ is a normalization factor. Clearly, for $V_0=0$ we obtain the exact solutions for the rigid box (\ref{eq29}). Using the trigonometric identity
\begin{equation}\label{eq76}
\tan^2(x)=\sec^2(x)-1
\end{equation}
we can easily obtain the energy eigenvalues for the Schr\"odinger equation
\begin{equation}\label{eq77}
\left[ -\frac{d^{2}}{dx^{2}}+V_{0}\sec^{2}\left(\frac{x}{a}\right)\right] \Psi
(x)={\cal E}_{\sec^2(x/a)}\Psi (x),\qquad {\Psi }\left(-\frac{\pi a}{2}\right) =\Psi\left(\frac{\pi a}{2}\right)=0.
\end{equation}%
where
\begin{equation}\label{eq78}
{\cal E}_{\sec^2(x/a)}=E_{\tan^2(x)}+V_0={1\over a^2}\left(n^2+(2n+1)\sqrt{a^2V_0 +1}+a^2V_0\right)
\end{equation}
Using a simple scaling argument $x\rightarrow a x$, we can write \eqref{eq77} as
\begin{equation}\label{eq79}
\left[ -\frac{d^{2}}{dx^{2}}+U_{0}\sec^{2}\left(x\right)\right] \Psi
(a x)=\mathfrak{E}\Psi (a x),\qquad {\Psi }\left(-\frac{\pi }{2}\right) =\Psi\left(\frac{\pi }{2}\right)=0,
\end{equation}%
where $U_0=a^2 V_{0}$ and $\mathfrak{E}=a^2{\cal E}_{\sec^2(x/a)}.$ 
Using the identity
\begin{equation}\label{eq80}
\sec(x+{\pi\over 2})=-\csc(x)
\end{equation}
we can write \eqref{eq79} as
\begin{equation}\label{eq81}
\left[ -\frac{d^{2}}{dx^{2}}+U_{0}\csc^{2}\left(x\right)\right] \psi=\mathfrak{E}~\psi,\qquad {\psi}\left(0\right) =\psi\left(\pi \right)=0,\qquad \mathfrak{E}=n^2+(2n+1)\sqrt{U_0+1}+U_0.
\end{equation}%

\section{Complex Cotangent potential well}
\noindent In this section we discuss a quasi-exact trigonometric potential model with complex coefficients \cite{com},  
\begin{equation}\label{eq82}
{V(x)=}\left\{ 
\begin{array}{ll}
iv\cot (\frac{x}{a}),& 0<x<a\pi,\quad v\geq 0,i=\sqrt{-1}\\ 
\infty,&x=0\text{ and }\pi a%
\end{array}%
\right. 
\end{equation}%
We show that the asymptotic iteration method can provides the necessary conditions under which the one dimensional Schr\"odinger equation 
\begin{equation}\label{eq83}
\left[ -\frac{d^{2}}{dx^{2}}+iv\cot\left(\frac{x}{a}\right)\right] \psi (x)=E\psi(x),\qquad {\psi }\left( x\rightarrow 0\right) ={\psi }%
\left( x\rightarrow a\pi \right) =0
\end{equation}
has exact (analytic) solutions.
To this end we assume that the wave function $\psi(x)$ takes the form
\begin{equation}\label{eq84}
\psi (x)=\sin \left(\frac{x}{a}\right)f(x),
\end{equation}%
which yields, up on substituting into equation \eqref{eq83}, the following differential equation for $f(x)$ 
\begin{equation}\label{eq85}
f^{\prime \prime }(x)=-\frac{2}{a}\cot\left(\frac{x}{a}\right)f^{\prime }(x)+\left( 
\frac{1}{a^{2}}-E+V_{0}\cot\left(\frac{x}{a}\right)\right) f(x)
\end{equation}%
wherein $V_{0} =iv$ the change of variable $y=\cot ({x}/{a})$ allows us to write (\ref{eq85}) as
\begin{equation}\label{eq86}
f^{\prime \prime }(y)=\left( \frac{1-a^{2}E+a^{2}V_{0}y}{(1+y^{2})^{2}}%
\right) f(y)
\end{equation}%
and finally, using the substitution
\begin{equation}\label{eq87}
f(y)=e^{-\beta~ \text{arccot}(y)}(1+y^{2})^{\alpha/2}g(y),\qquad -\infty<y<\infty
\end{equation}%
where $\alpha$ and $\beta$ are constants to be determine shortly, we obtain%
\begin{equation}\label{eq88}
g^{\prime \prime }(y)=-\frac{2(\alpha y+\beta)}{1+y^{2}}g^{\prime }(y)-\frac{\alpha(\alpha-1)y^2+(2\beta(\alpha-1)-V_0a^2)y+\alpha+Ea^2+\beta^2-1}{(1+y^{2})^{2}}g(y)
\end{equation}%
To solving this equation we note first that
\begin{equation}\label{eq89}
\beta= {a^2 V_0\over 2(\alpha-1)},\qquad \alpha\neq 1
\end{equation}
which yields
\begin{equation}\label{eq90}
g^{\prime \prime }(y)=-\frac{2(\alpha y+\beta)}{1+y^{2}}g^{\prime }(y)-\alpha(\alpha-1)\left(\frac{\frac{\alpha+a^2E-1+\beta^2}{\alpha(\alpha-1)}+y^2}{(1+y^{2})^{2}}\right)g(y)
\end{equation}%
Thus with
\begin{equation}\label{eq91}
a^2 E={(\alpha-1)^2-\beta^2}
\end{equation}
we obtain a more compact form of the second order linear differential equation for $g(x)$ as
\begin{equation}\label{eq92}
g^{\prime \prime }(y)=-\frac{2(\alpha y+\beta)}{1+y^{2}}g^{\prime }(y)-\frac{\alpha(\alpha-1)}{1+y^{2}}g(y)
\end{equation}
The asymptotic iteration method can then be applied with
\begin{equation}\label{eq93}
\left\{ \begin{array}{l}
\lambda_0=-\frac{2(\alpha y+\beta)}{1+y^{2}}\\ \\
s_0=-\frac{\alpha(\alpha-1)}{1+y^{2}}
       \end{array} \right.
\end{equation}
to yield, by means  of the termination condition \eqref{eq9}, the following condition for an exact solution  
$$\delta_n=0\Longrightarrow {\alpha^2(\alpha-1)\over\alpha+n}\prod_{k=1}^n(\alpha+k)^2=0\rightarrow \alpha=-n,\quad n=1,2,\dots\quad\text{for}\quad \alpha\neq 1.
$$ 
The (analytic) polynomial solutions are given, using \eqref{eq6}, explicitly as
\begin{align*}
n&=0\Longrightarrow g_0(y)=1\\
n&=1\Longrightarrow g_1(y)=-2y+2\beta\\
n&=2\Longrightarrow g_2(y)=3y^2-6\beta y+2\beta^2-1\\
n&=3\Longrightarrow g_3(y)=-3y^3+9\beta y^2-(6\beta^2-3)y+\beta^3-2\beta\\
n&=4\Longrightarrow g_4(y)=15y^4-60\beta y^3+30(2\beta^2-1)y^2-20\beta(\beta^2-2)x+2\beta^4-10\beta^2+3\\
\dots&=\dots
\end{align*}
The recurrence relation to generate these polynomial solution is given by
\begin{equation}\label{eq94}
g_{n+2}=\left({2(2n+1+2\alpha)(n+1+\alpha)y\over n+2\alpha}+{2\beta(2n+1+2\alpha)(\alpha-1)\over (n+2\alpha)(n+\alpha)}\right)g_{n+1}+
{4(n+1)(n+1+\alpha)(n^2+2\alpha n+\alpha^2+\beta^2)\over (n+2\alpha)(n+\alpha)}g_n
\end{equation}
initiated with
 \begin{equation}\label{eq95}
g_{0}=1,\qquad g_1=2\alpha y+2\beta
\end{equation}
In terms of the hypergeometric functions, the polynomial solutions can be written, are
\begin{equation}\label{eq96}
g_n(y)=C_n(-2i)^n\left(\alpha+i\beta\right)_n{}_2F_1\left(-n,n-1+2\alpha;\alpha+i\beta;{1\over 2}-{iy\over 2}\right),\qquad n=0,1,2,\dots
\end{equation}
The normalization condition $C_n$ can be computed using
\begin{align}\label{eq97}
\int_{-\infty}^\infty g_n(y)g_m(y) (1+y^2)^{\alpha-1}e^{2\beta\arctan(y)}dy&={(-1)^n4^{n+\alpha}(\alpha+\beta i)_n\pi \Gamma(1-2\alpha)\over \Gamma(-\alpha+\beta i+1)\Gamma(-\alpha-\beta i+1)}{(\alpha-\beta i)_n\over (2\alpha)_n}{(2\alpha +n-1)\over (2\alpha+2n-1)}~n!~\delta_{nm}
\end{align}
which is valid for $\alpha < \frac{1}{2}-n,~~n = 0,1,2,\dots,$ where $\delta_{nm}=0$ for $n\neq m$ and  $\delta_{nm}=1$ for $n= m$.

\section{Conclusion}
\noindent We have applied AIM to solve Schr\"odinger's equation with a number of confining trigonometric potentials. When a termination condition is satisfied, the method yields exact analytical solutions. In other cases approximations are found either by a  direct application of AIM or by using the method to find exact expressions for the terms in a perturbation expansion. In particular, our results lead to a new exactly solvable problem with a complex cotangent potential which is infinite outside the interval $\left( 0,\pi a\right) $.
\vskip0.1true in
\noindent The asymptotic iteration method can be applied to a wide variety of one-dimensional or spherically symmetric quantum potentials defined on infinite or semi-infinite domains. In the present work, the method is applied to a number of quantum (trigonometric) potentials defined on finite domains $[a,\,b]$, for which the method proves to be extremely effective for finding both analytical and approximate solutions. It is also clear from the present work that the method is applicable to periodic potentials and it is our intention to treat such problems in future work. 

\section{Acknowledgements}
\medskip
\noindent Partial financial support of this work under Grant Nos. GP3438 and GP249507 from the
Natural Sciences and Engineering Research Council of Canada
 is gratefully acknowledged by two of us (RLH and NS).

\end{document}